\documentclass[12pt]{iopart}

\usepackage{graphicx}
\usepackage{gensymb}

\begin{document}

\title[Method to make a single-step etch mask for 3D monolithic nanostructures] {Method to make a single-step etch mask for 3D monolithic nanostructures }  
\author{D A Grishina$^1$, C A M Harteveld$^1$, 
L A Woldering$^2$ and W L Vos$^1$}

\address{$^1$ Complex Photonic Systems (COPS), MESA+ Institute for Nanotechnology, University of Twente, PO Box 217, 7500 AE Enschede, The Netherlands}
\address{$^2$ Transducer Science and Technology (TST), MESA+ Institute for Nanotechnology, University of Twente, 7500 AE Enschede, The Netherlands}

\ead{d.grishina@utwente.nl}

\vspace{10pt}

\date{DG141104 \today}

\begin{abstract}

Current nanostructure fabrication by etching is usually limited to planar structures as they are defined by a planar mask. The realisation of three-dimensional (3D) nanostructures by etching requires technologies beyond planar masks. We present a method to fabricate a 3D mask that allows to etch three-dimensional monolithic nanostructures by using only CMOS-compatible processes. The mask is written in a hard-mask layer  that is deposited on two adjacent inclined surfaces of a Si wafer. By projecting in single step two different 2D patterns within one 3D mask on the two inclined surfaces, the mutual alignment between the patterns is ensured. Thereby after the mask pattern is defined, the etching of deep pores in two oblique directions yields a three-dimensional structure in Si. As a proof of concept we demonstrate 3D mask fabrication for three-dimensional diamond-like photonic band gap crystals in silicon. The fabricated crystals reveal a broad stop gap in optical reflectivity measurements. We propose how 3D nanostructures with five different Bravais lattices can be realised, namely cubic, tetragonal, orthorhombic, monoclinic, and hexagonal, and demonstrate a mask for a 3D hexagonal crystal. We also demonstrate the mask for a diamond-structure crystal with a 3D array of cavities. In general, the 2D patterns for the different surfaces can be completely independent and still be in perfect mutual alignment. Indeed, we observe an alignment accuracy of better than 3.0 nm between the 2D mask patterns  on the inclined surfaces, which permits one to etch well-defined monolithic 3D nanostructures.

\end{abstract}

%
%

\section{Introduction}

Progress in nanofabrication techniques is a key factor that enables a rapid growth of nanotechnology and its applications, because nanostructured materials exhibit unique complex behaviour. While many efforts are devoted to developing one-dimensional (1D) and two-dimensional (2D)  structures \cite{siphot,2Dstr,lukas}, opportunities of three-dimensional (3D) structures - such as 3D bandgap, sensing, opto-electronics and nano-electronics \cite{Rev.Fab,Arpin2010,ralf} - are less explored because of challenges in their realization. An important additional demand for 3D nanofabrication is to use only CMOS compatible techniques to allow for applications in micro and nano-electronics \cite{Judy2001,vlas10}. Additional requirements to 3D nanostructures are a high purity, and in case of nanophotonic structures, a high refractive index contrast and low roughness. 

In this work we concentrate on the fabrication of 3D nanophotonic structures and in particular 3D photonic crystals. 3D photonic crystals have already been realized using a large variety of techniques \cite{vlasov,lopez03,braun13,galis10,book_loc,FIB}. Firstly, one popular way to fabricate 3D nanostructures is the family of template-assisted methods \cite{templ1,templ2,pine07}. In these methods, one first assembles a template that is infiltrated with a high refractive index material, followed by removal of the template by calcination or etching. A large variety of templates has been demonstrated, such as artificial opals made from colloidal nanoparticles (usually polymer or silica), polymer photoresist structured by 3D holography \cite{campbell}, or resist structured by direct laser writing (DLW) \cite{sun1999,DLW_nature99,deubel}. The geometry of the structures fabricated with DLW is well defined and can be very complex. Nevertheless, subsequent inversion to a high refractive index material introduces undesirable yet unavoidable impurities, roughness, and undesired absorption \cite{rough}. Secondly, impressive structures have been created using layer-by-layer methods \cite{leon29,Noda2004}. Remarkable results have been reported on the fabrication of woodpile photonic crystal structures \cite{noda_layers,noda_science}. The main difficulty of layer-by-layer fabrication is the alignment between layers: each layer has to be carefully aligned with respect to the previous one. Here we call such an alignment a planar alignment since the layers that are being aligned are in parallel planes. In CMOS industry the parameter characterizing the planar alignment between two layers is called overlay \cite{book}. For the fabrication of an \textit{N}-layer thick structure \textit{N-1} alignment steps are needed. In order to avoid a large number of alignment steps a third class of fabrication methods has been proposed, where 3D nanostructures are created by consecutively patterning an etch mask on only two adjacent oblique wafer surfaces with \emph{only} a single alignment step, followed by deep etching \cite{tjerk2011,deviation,grooves}. It is important to highlight that in this case the alignment is not planar anymore, as the masks that are aligned with respect to each other are lying in oblique planes. For the diamond-like photonic crystal under study, the out-of-plane alignment should be better than 50 nm \cite{deviation}. In practice an out-of-plane alignment of 15 nm was reached \cite{tjerk2011}. Nevertheless, even one such out-of-plane alignment step introduces significant complexity in the fabrication procedure, introducing deviations from perfect alignment and extending the time needed for fabrication.

\begin{figure}[h]
 \includegraphics[totalheight=8cm]{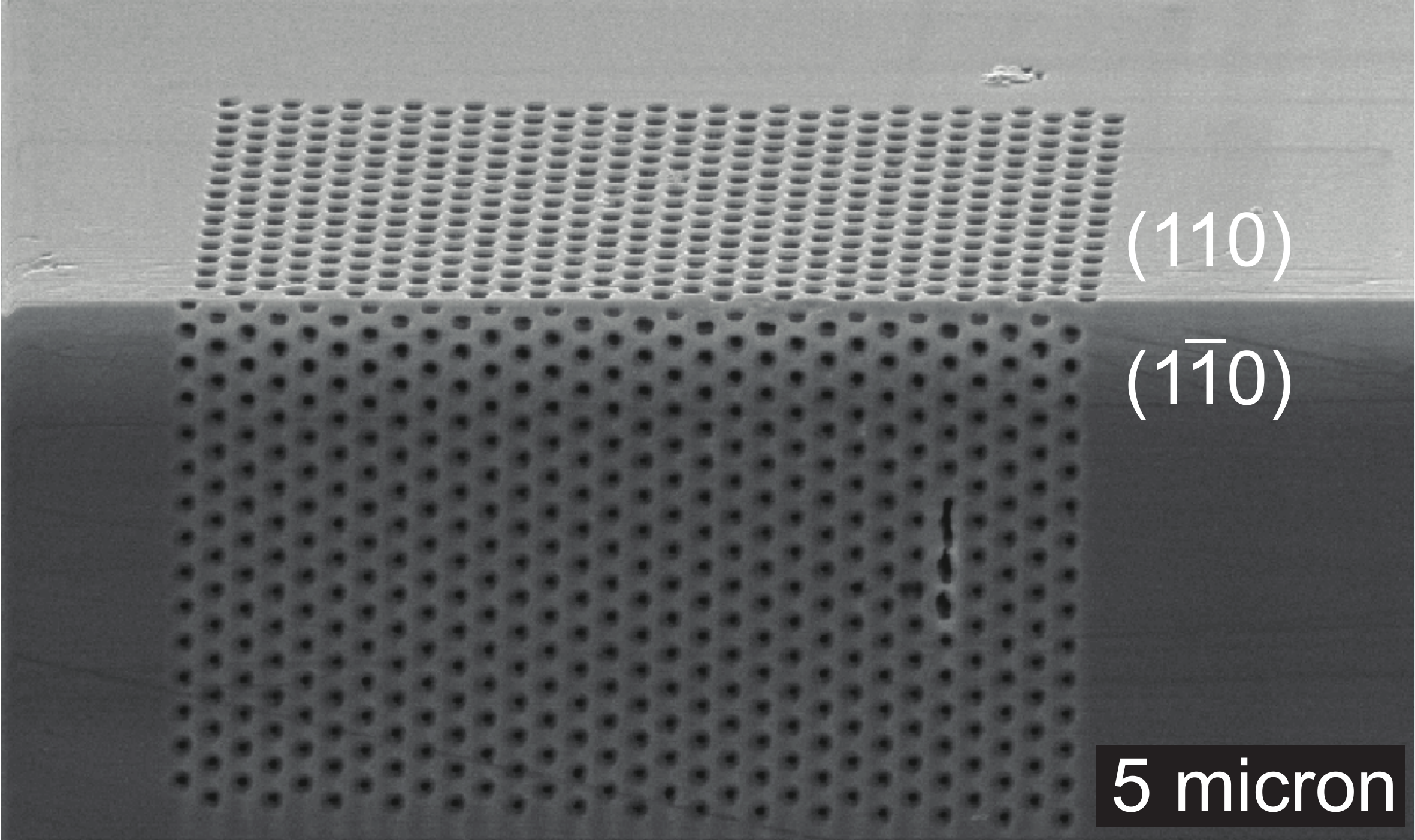}
 \caption{\label{FIG1} SEM image of a monolithic 3D photonic band gap crystal fabricated in Si using a single step etch mask. The crystal has the inverse woodpile structure with a cubic diamond-like symmetry that consists of two sets of perpendicular pores. The top surface in the image is the (110) crystal plane and the perpendicular surface at the bottom is the $(1\bar{1}0)$ crystal plane. Scale bar is shown in the image.}%
 \end{figure}

In this paper we present a new fabrication method for monolithic 3D nanostructures. The key step is to define and make a single step etch mask on two inclined surfaces \emph{simultaneously} with alignment ensured at the design stage. As an example for a single step etch mask, we realise a pattern that yields a (110) plane of a cubic inverse woodpile structure on one surface and $(1\bar{1}0)$ plane on a perpendicular adjacent surface. We describe the fabrication process of such a so-called ``3D mask'', as well as the subsequent etching process to fabricate 3D photonic band gap crystals with diamond-like inverse woodpile structure (see Figure \ref{FIG1}). We characterize the ensured alignment at the design stage by carefully characterization of the realised structures on the inclined surfaces. We study nanophotonic behaviour of the fabricated crystal by reflectivity measurements. We also discuss other 3D structures that are feasible using our method such as disordered structures and different 3D Bravais lattices; as a proof of principle, we demonstrate a 3D mask for a hexagonal 3D nanostructure, and a 3D cavity array.

\section{Fabrication process for the 3D single-step etch mask with built-in alignment}
    
\begin{figure}[h]
 \includegraphics[totalheight=15cm]{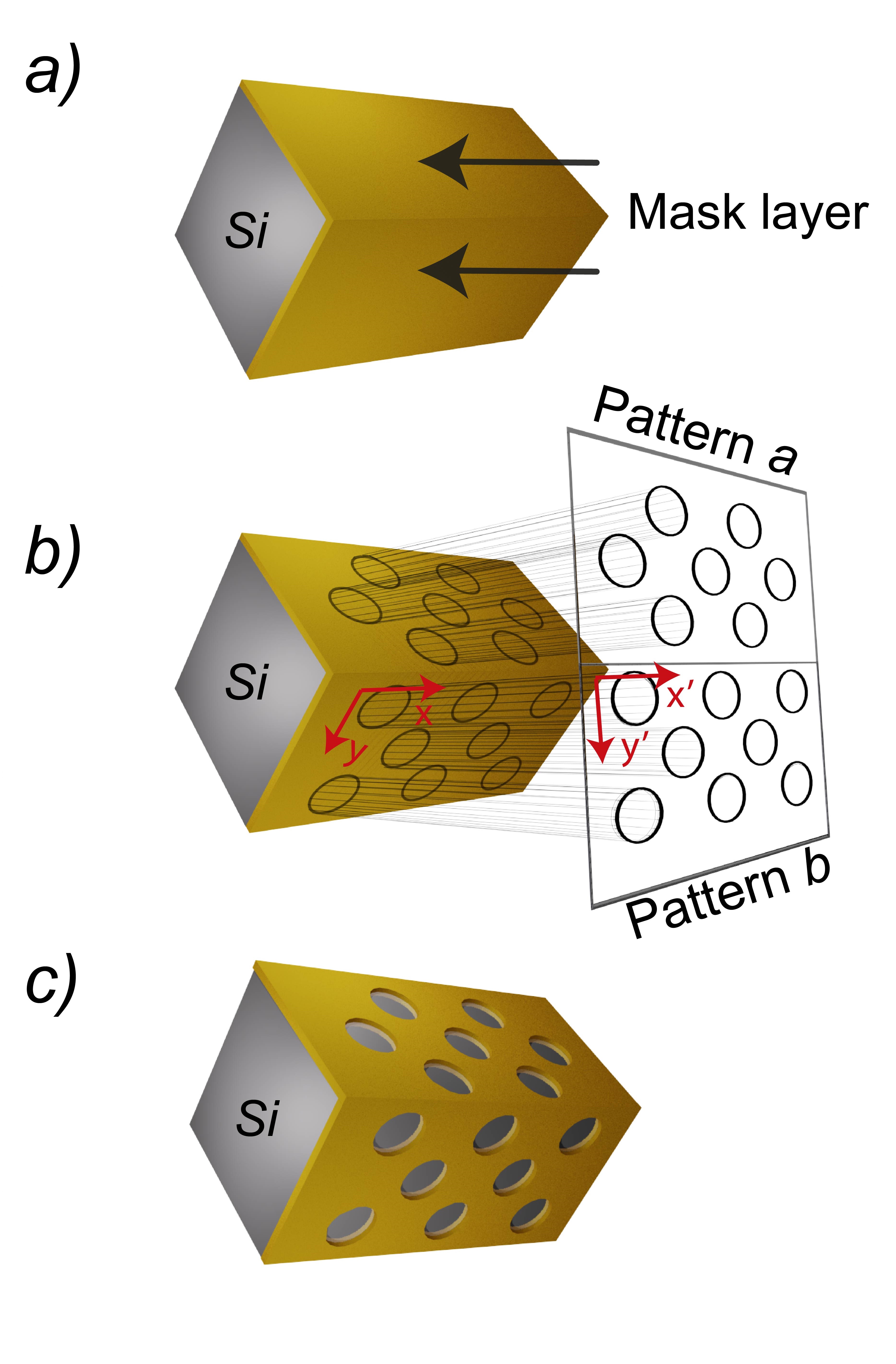}
 \caption{\label{FIG2}Scheme for the  single step etch mask fabrication on two perpendicular surfaces. a) Deposition of a hard mask layer on two inclined surfaces of a Si wafer. b) Patterning of mask layer in one step on both surfaces - projection of single 2D mask on 3D surface. Patterns for both perpendicular surfaces are written in one projection and therefore alignment is ensured. c) Apertures are opened in the mask layer to obtain an etch mask for two intersecting 2D structures that yield the desired 3D structure.  }%
 \end{figure}

  The generic scheme to fabricate a 3D etch mask on two inclined surfaces is shown in Figure \ref{FIG2}. First, the hard mask material that serves as an etch stop is deposited on two inclined surfaces, see Figure \ref{FIG2}(a). Next, the desired pattern is projected onto the oblique surfaces, as shown in Figure \ref{FIG2}(b). The projection is made from a side such that both inclined surfaces can be reached. The projected pattern consists of two parts: pattern \textit{a}, designed for one surface and pattern \textit{b}, designed for the second surface. We emphasize that our method allows for a complete freedom to independently design the two patterns \textit{a} and \textit{b}.  In Figure \ref{FIG2}(b) the two patterns are designed to be similar. The two patterns are written in one projection and are therefore by design in perfect mutual alignment. Mask apertures for subsequent etching are opened during the third step of the mask fabrication process (Figure \ref{FIG2}(c)). 
  
  The fact that the pattern is projected on a non-normal surface must be taken into account in the pattern design. The pattern is designed in such a way that after projection on the inclined surfaces it yields the desired structure. As is shown in Figure \ref{FIG2}(b), the $x'$ and $y'$ coordinates in the design plane differ from the \textit{x} and \textit{y} coordinates on the sample surface.

  \begin{figure}[h]
 \includegraphics[totalheight=7cm]{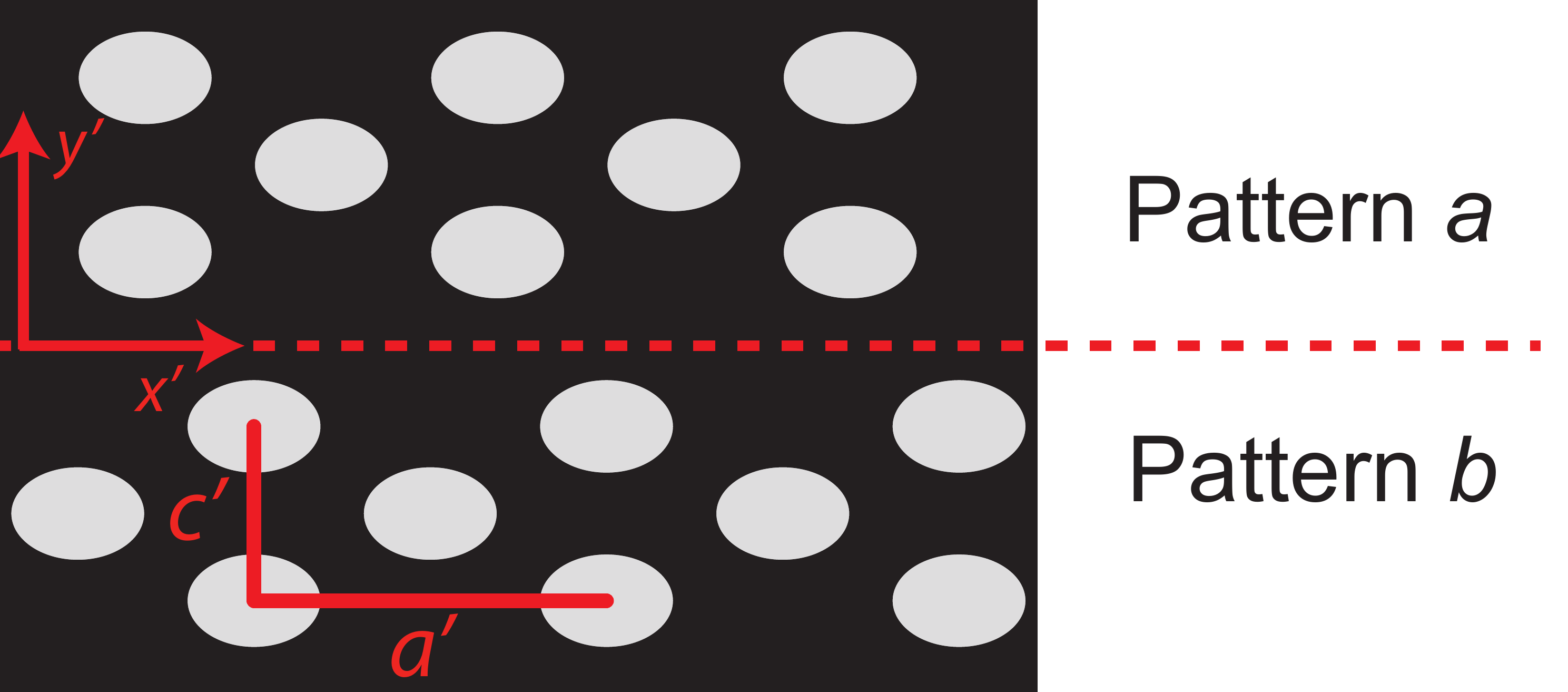}
 \caption{\label{FIG3}  Geometry of the pattern that is projected on two inclined surfaces. The structure is designed to consist of two centred rectangular arrays shifted with respect to each other by $x'=c'/4$ such that after being projected on two {45\degree}   inclined surfaces it gives the (110) and $(1\bar{1}0)$ faces of a cubic inverse woodpile photonic crystal with lattice parameters  \textit{a} = $\sin{45\degree}\times a' $ and $c=c'$. The basic building blocks are designed to be elliptical, in order to yield circles on the mask after projection. The dashed line delimits the top part of the pattern (pattern \textit{a}) that is projected on one surface of the wafer and the bottom part (pattern \textit{b}) that is projected on the second surface.  }%
 \end{figure}

We demonstrate our fabrication process using as an example a 3D cubic diamond-like photonic band gap crystal made from silicon. Due to its physical properties silicon plays an important role both in optics and electronics. As a material that is widely used in research and manufacturing, silicon is widely available, cheap, and has a very high purity. Among the diamond-like photonic band gap structures \cite{Mald2004} we chose the inverse-woodpile photonic crystal. Inverse-woodpile photonic crystals deserve particular attention in view of the broad band gap with relative width ($\Delta \omega_{gap}/\omega_{gap}$) of more than 25\% \cite{deviation,ho1994,Hill}. These crystals consist of two mutually perpendicular rectangular arrays of cylindrical pores etched in a high refractive index material. Conceptually the fabrication of an inverse-woodpile structure is easy, although its fabrication remains a challenge due to the required precise alignment of the perpendicular sets of pores \cite{deviation,adv.mat12,tjerk2011}.

We start the fabrication procedure from a single crystalline Si wafer. We fabricated the 3D etch mask on two polished adjacent perpendicular surfaces of the wafer. In the first step we deposited a 50 nm thick Cr layer that serves as a hard mask material on two adjacent surfaces of a wafer (Figure \ref{FIG2}(a)). We choose Cr as a hard mask material due to its sustainability to $SH_6$ etching \cite{etch}, but other possible mask materials such as SiN or SiC are also compatible with our method. The deposition of Cr is done in a home-built sputtering machine and takes around 4 minutes for a 50 nm thick layer. 
 
The patterning of the etch mask can be performed using many types of lithography \cite{Phys_today} such as e-beam, deep UV (DUV) step-and-scan, or nanoimprint. In our case we had focused ion beam (FIB) milling equipment at our disposition to project the mask and open the apertures. We placed a sample in a FEI Nova 600 Nanolab FIB chamber under {45\degree} angle with respect to the ion beam gun so that both adjacent surfaces of the wafer can be reached (Figure \ref{FIG2}b). 

The design of a single pattern that is projected on two adjacent surfaces consists of two parts as shown in Figure \ref{FIG3}: pattern \textit{a} intended for one surface and pattern \textit{b} intended for the second surface. 
 In our case the surfaces are orthogonal to each other and aligned at {45\degree} angle with respect to the ion beam gun in \textit{y} direction (Figure \ref{FIG2}(b)). In the pattern design it is taken into account that projection of the pattern is made under an angle $\theta=45\degree$ to the surfaces, meaning that $x'$ and $y'$ coordinates in the pattern design are related to the $x$ and $y$ coordinates on a surface of fabricated mask as following: $x'=x$ and $y'=y/\sin{\theta}$. In order to form a cubic diamond-like structure inside the silicon wafer we patterned each surface of the wafer with a centred rectangular array of holes with lattice parameters \textit{a} and \textit{c}, where $\frac{\textit{a}}{\textit{c}}=\sqrt{2}$ to fulfil the criterion for a cubic crystal. Arrays of holes for two surfaces are shifted by $c/4$ in the \textit{x}-direction. The pattern is shown in Figure \ref{FIG3} and yields (110) and $(1\bar{1}0)$ crystal surfaces on a surface of a wafer. The mask pattern is projected on both surfaces in one step. Since the two patterns intended for different surfaces are contained in one image, the alignment between them is ensured. A patterning of a 3D mask consisting of two arrays of  30 by 30 holes each takes 7 minutes.

After the etch mask is created, the next step is to etch deep pores inside Si through the openings. Etching can be done using a variety of techniques \cite{etch_rew} such as  reactive ion etching (RIE) \cite{nanot19}, cryogenic etching, wet etching, photoelectrochemical etching \cite{shill01,holl}, or other types of etching depending on a desired structure ans mask material. Nanopores are first etched in one direction, then the sample is rotated by {90\degree} and pores are etched in the second perpendicular direction. We etched deep nanopores using deep reactive ion etcher (DRIE) Adixen AMS 100SE as described in reference \cite{nanot19} with an additional oxygen plasma cleaning step that is performed after etching each set of pores. The cleaning step is needed to remove the protective polymer layer remaining after the etching process into the front surface of the sample.

\section{Results and discussion}
\subsection{Fabricated structure characterization}

\begin{figure}[h!]
 \includegraphics[totalheight=18cm]{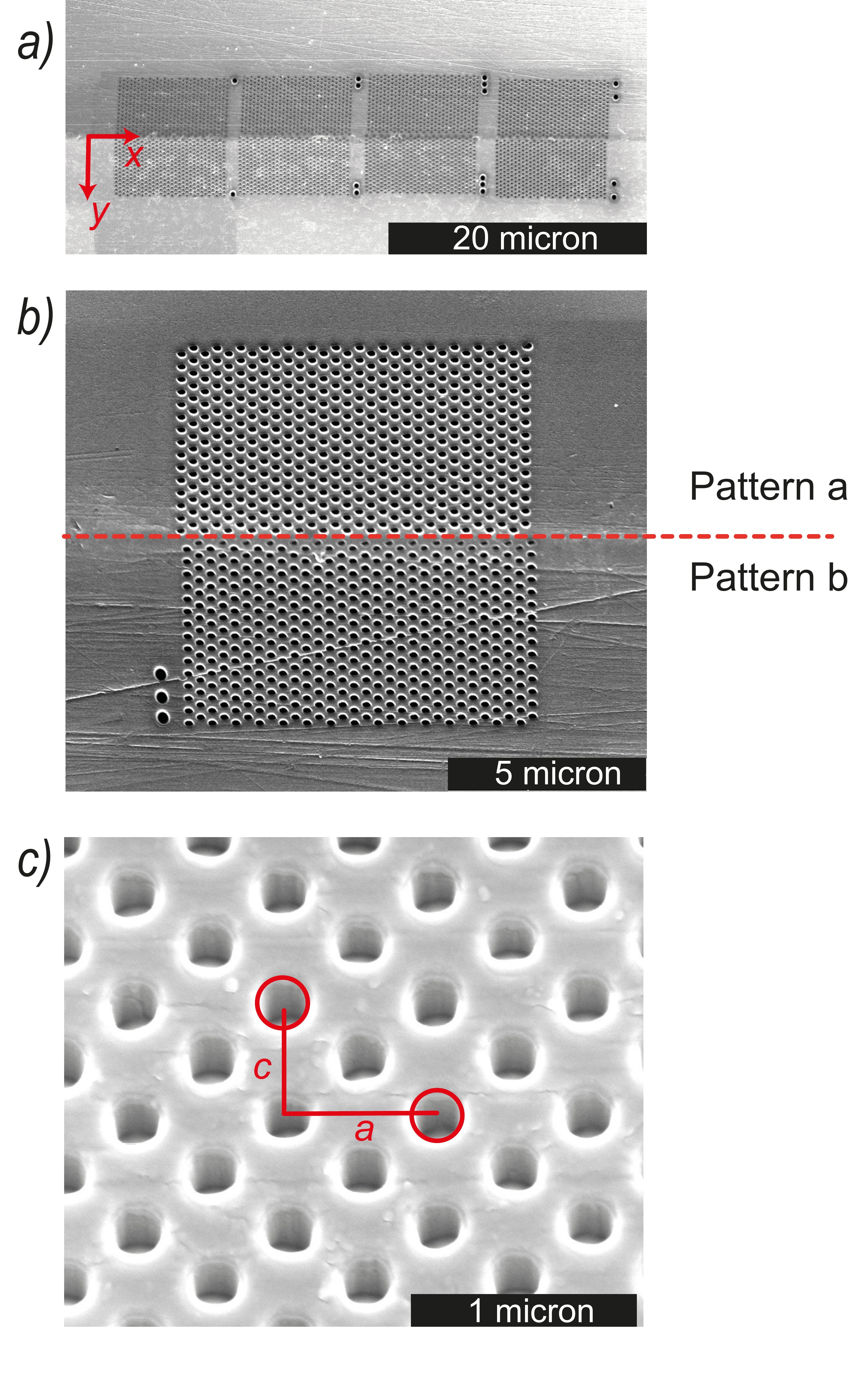}
 \caption{\label{FIG4} a) Overview of a Si wafer with four 3D etch masks milled in one step. The coordinates \textit{(x,y)} are indicated. b) Side view on one of the mask patterns. The dashed line in the middle indicates the 90 degree edge of the Si wafer. c) Zoom-in on one surface of a mask pattern. \textit{a} and \textit{c} are lattice parameters with $\frac{a}{c} = \sqrt{2}$ . Scale bars are shown in each image.  }%
 \end{figure}

The result of patterning a 3D etch mask in a single step is shown in Figure \ref{FIG4}. Figure \ref{FIG4}(a) shows four 3D masks in a row on the edge of a Si wafer.  Each of the 3D masks was written in one step and consists of (110) and $(1\bar{1}0)$ crystal planes of the cubic inverse woodpile crystal. The number of 3D mask structures written along the edge of a Si wafer is only limited by the size of the wafer. The maximum width in the \textit{x}-direction of a single structure that is written in one step is determined by the horizontal field of view of the lithography tool which is in our case a FIB setup. In this study the horizontal field of view was 12.8$\mu m$ as set by the magnification of 10000$\times$. By decreasing the magnification or stitching the fields of view, it is possible to increase the size of a continuous structure in the \textit{x} direction. Thus, we may effectively consider the four closely spaced 3D patterns in Figure \ref{FIG4}(a) as one large $L_{x}=40{\mu}m$  sized nanostructure. The size in the \textit{y}-direction is limited by the depth of focus of the tool and can be extended by performing for example multiple milling runs at different depths. 

Figure \ref{FIG4}(b) shows one complete mask patterned on both wafer surfaces. Above the dashed red line there is a surface that contains pattern \textit{a} corresponding to a (110) crystal plane of the targeted photonic crystal. Below the dashed line there is a perpendicular surface that contains pattern \textit{b} corresponding to a $(1\bar{1}0)$ crystal surface. The design for this mask shown in Figure \ref{FIG3} is such that centres of the apertures in pattern \textit{b} are shifted by $\Delta x=a/4$ to be exactly in the middle between apertures of pattern \textit{a}. The lower surface of the wafer has some deep lines that are the result of manual polishing and which slightly reduce the quality of the mask layer on that side. Fortunately, the effect of such lines was found to be insignificant in subsequent processing, although it may introduce optical scattering. 

\begin{figure}[h!]
 \includegraphics[totalheight=17cm]{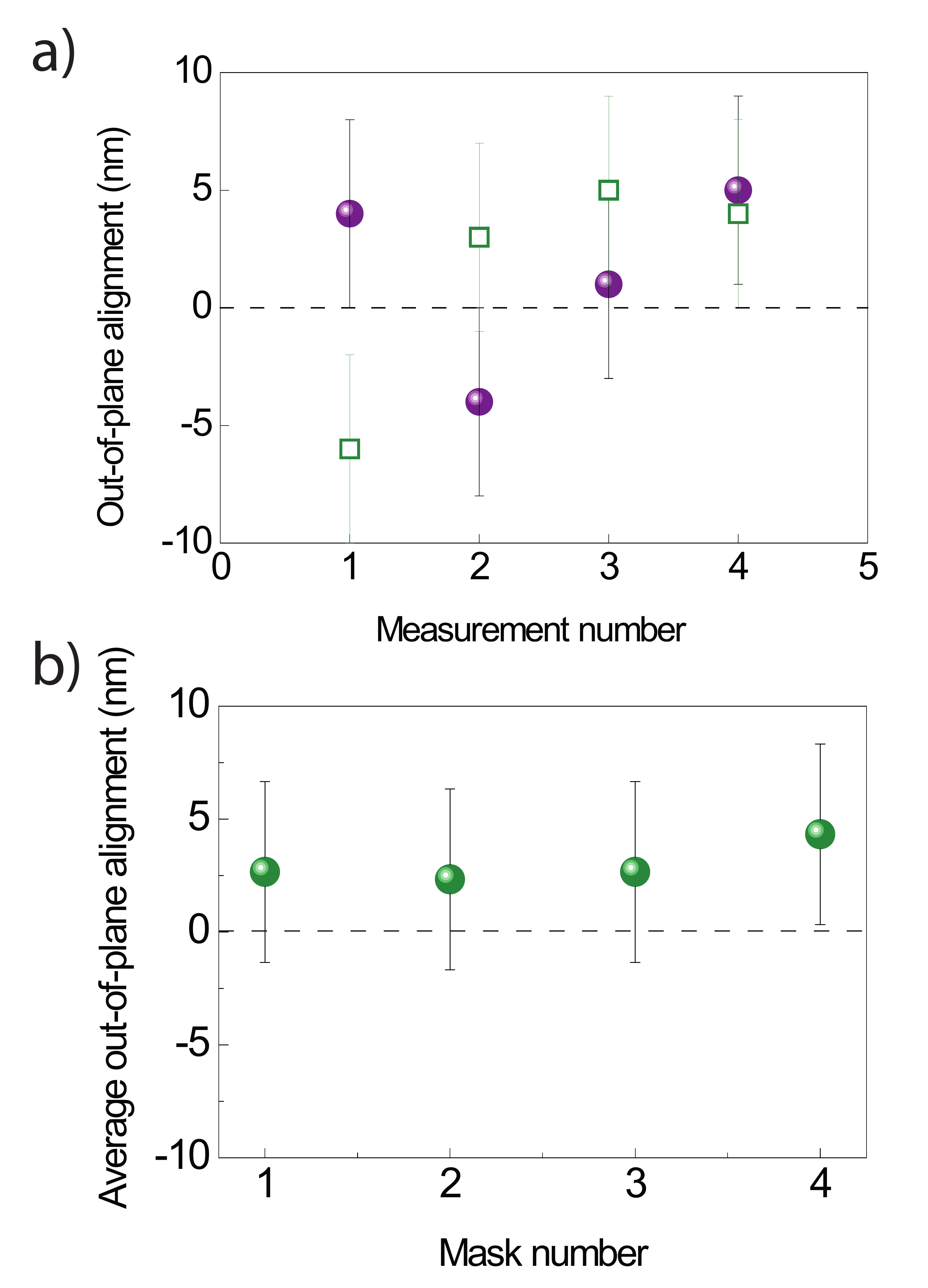}
 \caption{\label{FIG5} (a) The alignment between patterns on an oblique surfaces in a mask for a hexagonal structure (\textit{cf} Figure \ref{FIG9}). The out-of-plane alignment is determined as a deviation of a relative position of two apertures located on an oblique surfaces from the design.  Each measurement is taken for a pair of apertures on oblique surfaces within one mask. Circles show measurements taken for apertures close to the edge~(in pattern \textit{a} $y=1.42~\mu$m and in pattern \textit{b} $y=-1~\mu$m), squares are for apertures further from the edge(in pattern \textit{a} $y=1.42~\mu$m and in pattern \textit{b} $y=-5.42~\mu$m). (b) Averaged out-of-plane alignment over several pairs of apertures on obliqued surfaces. Each measurement is taken for a different mask. In both figures error bars represent the resolution of SEM. }%
 \end{figure}
 
From SEM images as in Figure \ref{FIG4}(b), we characterize the out-of-plane alignment, in other words, the alignment between patterns \textit{a} and \textit{b}. To characterize the out-of-plane alignment we take cross-sections in \textit{x}-direction through a row of apertures. First we take a cross-section in pattern \textit{a}, then in pattern \textit{b} and compare the relative position of apertures between the different patterns to the design. In Figure \ref{FIG5}(a) we have collected two sets of out-of-plane alignment data for different rows of apertures from Figure \ref{FIG9}. In the $1^{st}$ set indicated as circles in Figure \ref{FIG5}(a) the row in pattern \textit{a} is  located at a distance $y=$1.4 $\mu$m from the edge, while the row in pattern \textit{b} is located at a distance $y=-1~\mu$m from the edge. From the data we conclude that the apertures are aligned within the error bar of SEM accuracy. To verify alignment over further distances from the edge, in the $2^{nd}$ set of alignment data, shown as squares in Figure \ref{FIG5}(a), the row in pattern \textit{b} is located at distance $y=-5.42~\mu$m, while the row in pattern \textit{a} is the same. We see that the out-of-plane alignment stays within the resolution of SEM for both sets of data, independent of the distance from the edge. We find the deviation from the designed structure (Figure \ref{FIG3}) to be at most 5 nm. Figure \ref{FIG5}(b) shows an averaged out-of-plane alignment over several pairs of apertures for different 3D mask nanostructures. It is seen that the average out-of-plane alignment varies between $\Delta x$=2.3 nm and $\Delta x$=4.3 nm with a mean of 3.0 nm. 
The error bars in both Figures \ref{FIG5} (a) and \ref{FIG5}(b) indicate the typical accuracy of $\pm$4 nm of the scanning electron microscope. Thus, we conclude that the out-of-plane alignment data are mostly determined by the SEM error and are better than 3.0 nm. Therefore, the out-of-plane alignment for individual apertures is consistent with zero deviation, in agreement with the starting point of our 3D mask method that the two oblique pattern have built-in mutual alignment. 

Figure \ref{FIG4}c shows a zoom-in to the front surface of a wafer. We see that the elliptical apertures in the pattern design (see Figure \ref{FIG3}) have been correctly projected to become circular apertures with the diameter of 273 nm on the sample surface. From Figures \ref{FIG4} and \ref{FIG5} we conclude that we have successfully fabricated a desired 3D mask structure on two inclined surfaces. We emphasize that the alignment between the patterns on two inclined planes is ensured at the design stage since both patterns are written in one projection and is within the resolution of SEM.

\begin{figure}[h!]
 \includegraphics[totalheight=18cm]{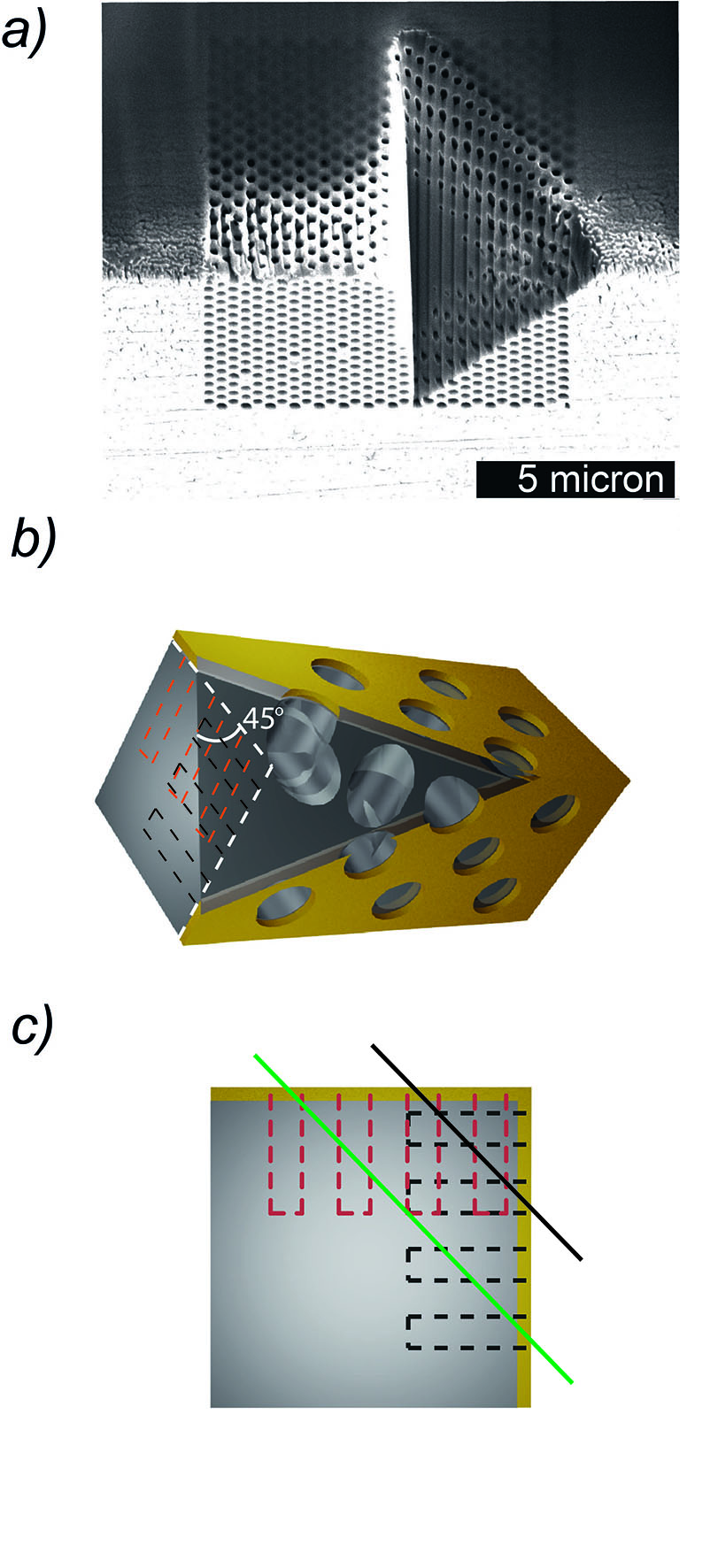}
 \caption{\label{FIG6}   (a) The fabricated 3D diamond-like photonic crystal that was opened up by focused ion beam milling. Scale bar is shown in the picture. (b) Schematic representation of how a cross-section is milled on a fabricated sample. The cut is made at  {45\degree} to the side of silicon bar. Dashed lines indicate the pore geometry. (c) Schematic cross-section that illustrates how limited pore depth appears on the milled structure. The black line shows where pores overlap with each other and the green line shows the depth beyond which the arrays of pores are two dimensional. (colour online) }%
 \end{figure}
 
 After etching of deep pores in silicon in two perpendicular directions, we have sacrificed one crystal in order to view the internal structure of the sample by milling it with a focused ion beam. Figure \ref{FIG6}(b) shows a schematic representation of the cut and a surface plane that is open for viewing. The cut was made under {45\degree} angle to both surfaces. Since the pore depth is finite, starting at a certain depth in the structure the perpendicular pores will not overlap inside the crystal. Pores that are far from the edge of a wafer are not deep enough in order to reach the corresponding perpendicular pore, see green line in Figure \ref{FIG6}(c). Therefore we expect to see a region inside the crystal closer to the edge where pores overlap and form a 3D structure. Further from the edge we expect a region where the structure will be two dimensional. The SEM image of a crystal cross section is shown in Figure \ref{FIG6}(a). We see that using a single step etch mask deep pores were successfully etched in silicon. Pores etched in both perpendicular directions form a 3D diamond-like structure in the bulk. Far from the surface of the Si wafer the pores are not overlapping. In the present case, the depth of the pores is determined to be 4 $\mu$m deep with an aspect ratio (depth to width ratio) of 14. The size of the fabricated photonic crystal is limited by the depth of the pores in the silicon \cite{nanot19}. We conclude from the cross-section that single step etch mask with ensured alignment allows us to fabricate a 3D monolithic nanostructure in bulk silicon.

\subsection{Nanophotonic behaviour}

\begin{figure}[h!]
 \includegraphics[totalheight=10cm]{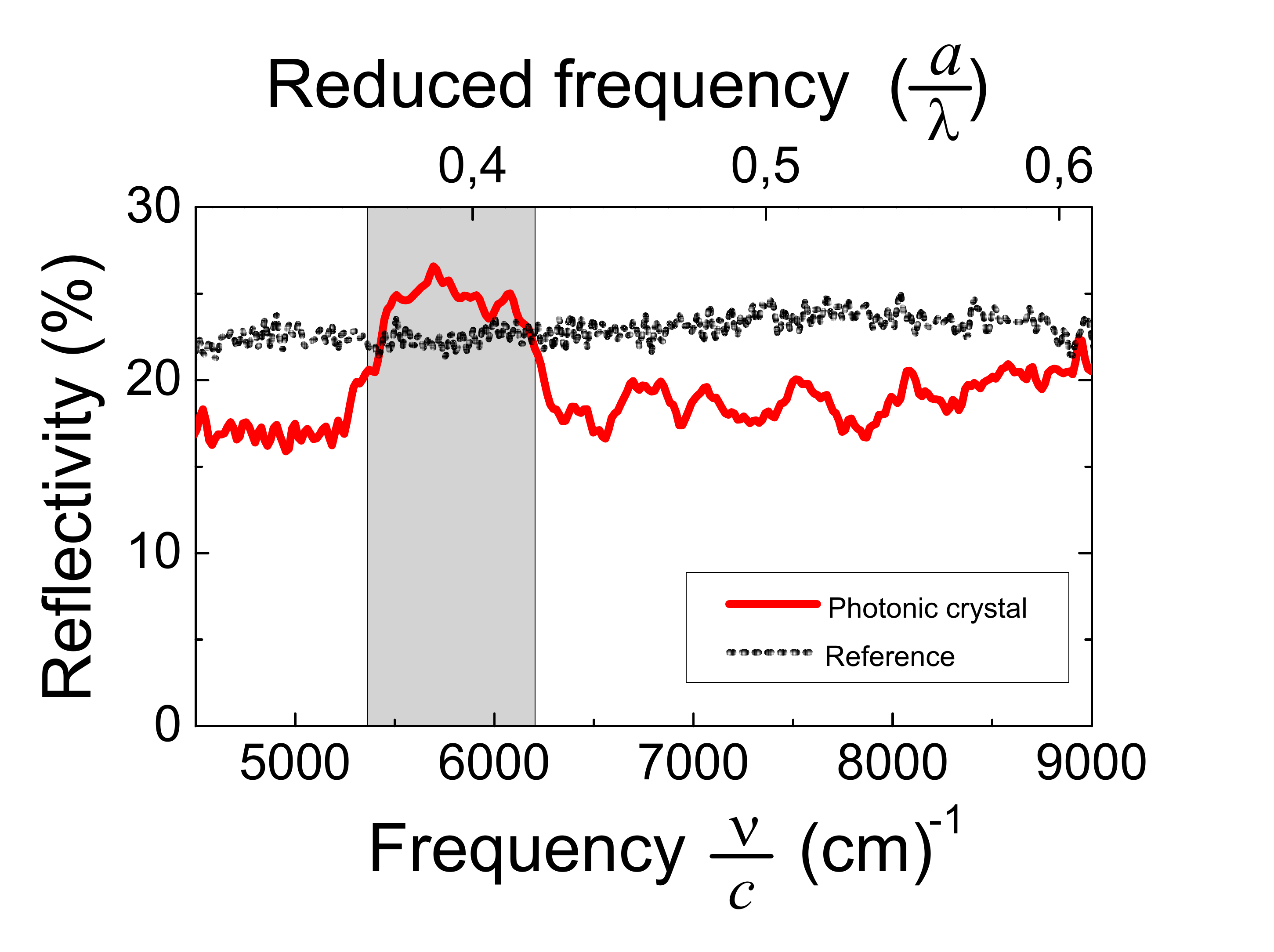}
 \caption{\label{FIG7} Reflectivity measurements on a fabricated 3D photonic crystal and a non-photonic reference. The grey bar indicates the stop gap calculated for a fabricated sample with lattice parameters taken from SEM images: $\frac{r}{a}$=0.16 and \textit{a}=679\textit{nm}.    }%
 \end{figure}

To verify the nanophotonic behaviour of the fabricated inverse-woodpile photonic band gap crystal, we have performed reflectivity measurements perpendicular to the (110) surface of a crystal. We used a method described in Ref. \cite{huisman}; in brief, a Fianium supercontinuum source was used as a broad band light source. The laser beam was focused on a sample surface with a reflective objective with a numerical aperture NA=0.65. Spectral measurements were performed using Biorad 600 Fourier-transform infra-red (FTIR) spectrometer with a resolution of 16 cm $^{-1}$. Figure \ref{FIG7} shows the measured reflectivity calibrated to a gold mirror. The experimental spectrum reveals a broad reflectivity peak from 5246 cm$^{-1}$ to 6341 cm$^{-1}$. The maximum reflectivity of 26\% is currently limited by finite size of a crystal, surface roughness and a relatively large beam size in the optical setup. The beam size during the experiment was close to 6 $\mu$m that is larger than a crystal size. Taking into account that crystals are necessarily located at the edge of the silicon wafer, in case when the beam size is larger than a crystal size, significant amount of laser light shines into the air resulting in low reflectivity value. The collected reflectivity spectrum averages over the illuminated area, therefore in case when the beam size is larger that a crystal size, reflection from a bulk silicon gives rise to the reflectivity value outside the stop gap and therefore decreases its maximum value. Nevertheless, the reflectivity peak indicates nanophotonic behaviour of a photonic crystal fabricated using a single step etch mask. The expected stop gap in $\Gamma$-Z direction according to plane wave band structure calculations in reference\cite{huisman} is situated from 5362 cm $^{-1}$ to 6204 cm $^{-1}$. Parameters for the stop gap calculations were taken from SEM images of the fabricated sample ($\frac{r}{a}$=0.16 and \textit{a}=679\textit{nm}).  Therefore, the observed stop band is in good agreement with the calculated stop gap for an infinite 3D photonic crystal. From observed nanophotonic behaviour of our sample we conclude that the functional photonic structure has been successfully fabricated using a 3D single step etch mask. 

\subsection{Structures feasible for fabrication}

\begin{figure}[h!]
 \includegraphics[totalheight=20cm]{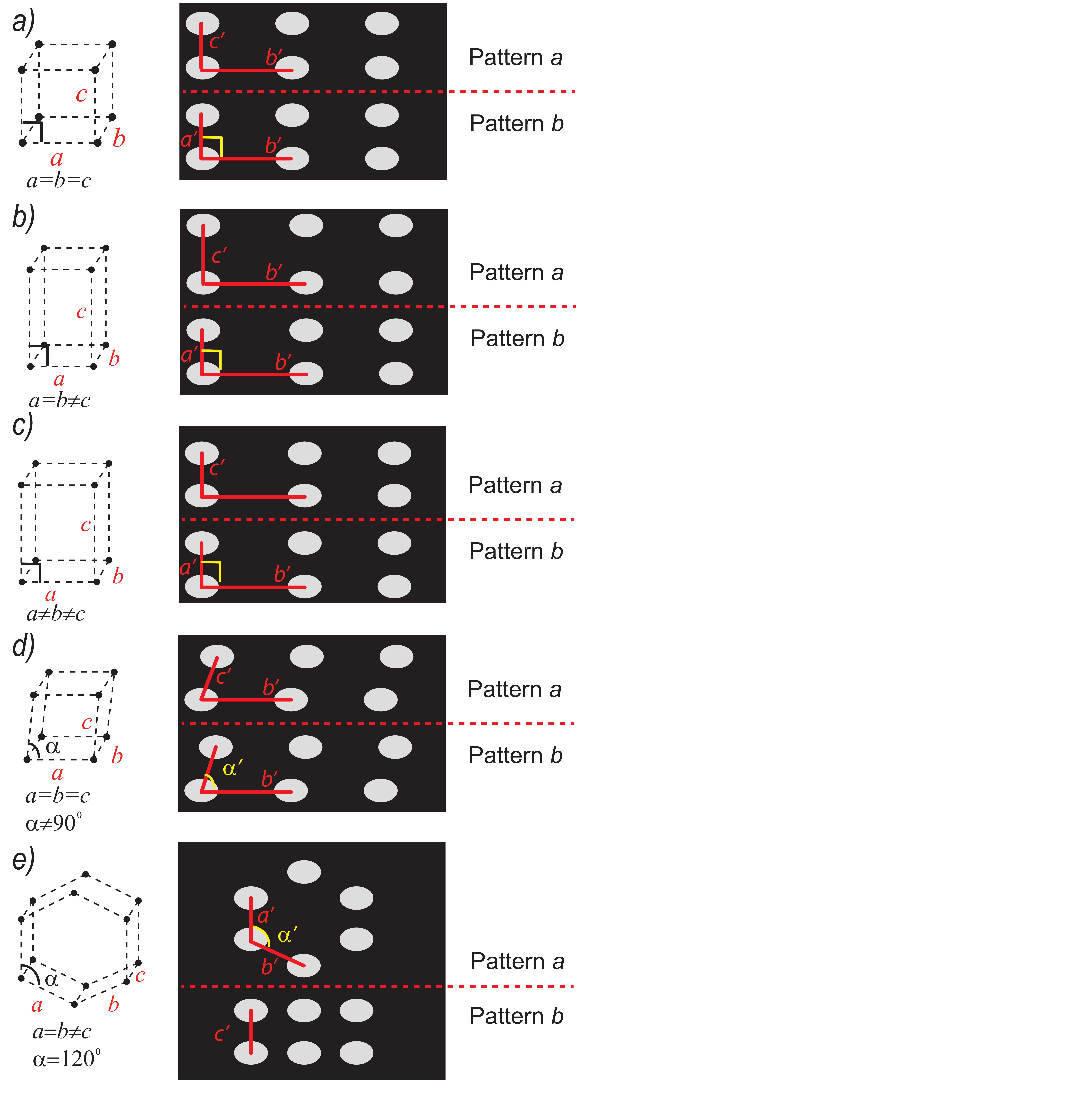}
 \caption{\label{FIG8}Pattern designs for different lattice systems fabrication. The 3D scheme of the structure is shown on the left and corresponding pattern for etch mask is on the right. (a) Cubic (b) Tetragonal (c) Orthorhombic (d) Monoclinic (e) Hexagonal   }%
 \end{figure}

\begin{figure}[h!]
 \includegraphics[totalheight=10cm]{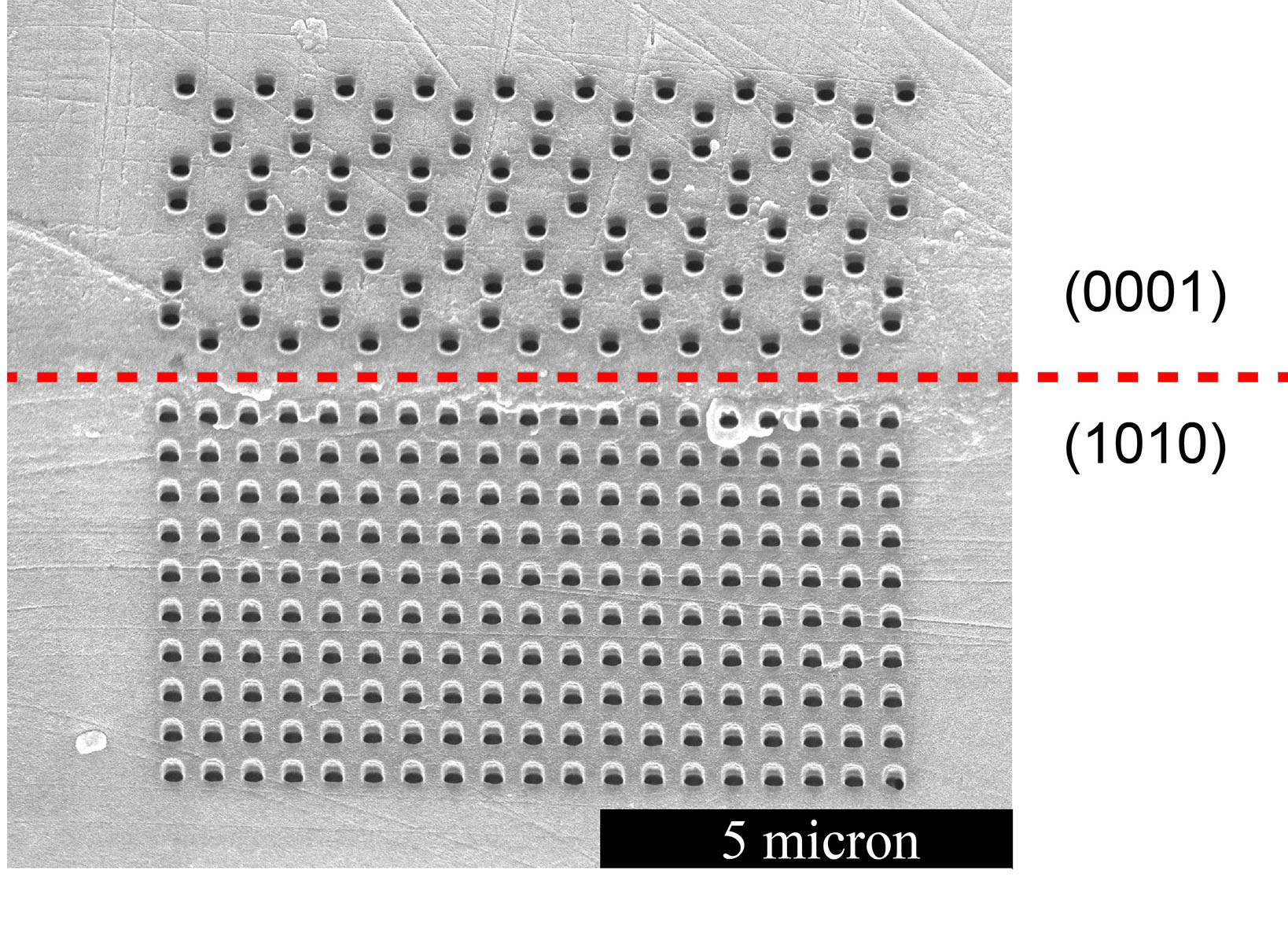}
 \caption{\label{FIG9} SEM image of a mask for hexagonal 3D crystal fabrication. The dashed line shows the $90\degree$ edge of a wafer.  Top side is a (0001) crystal plane of hexagonal structure and the bottom side is a (1010) crystal plane. The scale bar is shown on the picture. }%
 \end{figure}

There is a large variety of structures that can be fabricated using the presented technique. In general mask patterns projected in one step on inclined surfaces have an arbitrary structure and still be in perfect alignment with respect to each other. We consider 3D Bravais lattices that can be created with the described fabrication process. Using different pattern designs, it is possible to achieve structures with five Bravais lattice systems: cubic, monoclinic, orthorhombic, tetragonal and hexagonal. In Figure \ref{FIG8}(a) on a left side is shown the simple cubic lattice. On the right side there is a pattern design for the fabrication of a simple cubic structure, where all lattice parameters are equal (\textit{a}=\textit{b}=\textit{c}). Taking into account that the pattern is projected on a $\theta={45\degree}$ inclined surfaces, in the design  ${a'} =a/\sin{45\degree}$, $c' =c/ \sin{45\degree}$ and $b'=b$. This design consist of two rectangular arrays of apertures aligned with no shift with respect to each other. For completeness we note that this design for a cubic structure is not unique, as many different designs are feasible. In particular, the mask design presented in Figure \ref{FIG3} confirms that there are multiple possible designs for cubic structure, notable depending on the type of cubic symmetry (simple versus face centred). Figure \ref{FIG8}(b) shows a simple tetragonal structure on the left side and the design pattern for realisation of such structure on two perpendicular surfaces on the right side. In case of tetragonal structure \textit{a}=\textit{b}$\neq$\textit{c}, therefore in the design pattern $c'\neq a'$. That means that the complete mask pattern for a tetragonal structure consist of two different rectangular arrays of apertures, one with lattice parameters $a'$ and $b'$ and other one with parameters $c'$ and $b'$. Figure \ref{FIG8}(c) shows the orthorhombic structure on the left and the corresponding design pattern on the right. For an orthorhombic structure \textit{a}$\neq$\textit{b}$\neq$\textit{c}, which means that in the design pattern all lattice parameters are different. Hence the complete design consist of two different rectangular arrays of apertures with parameters $a'$,$b'$ and $c'$,$b'$. In Figure \ref{FIG8}(d) on the left side the monoclinic structure is shown, where \textit{a}=\textit{b}=\textit{c} and the angle between two lattice vectors is $\alpha\neq 90\degree$. In the pattern design this means that the rows of apertures are shifted with respect to each other so that the angle between vectors $a'$,$b'$ and $c'$,$b'$ is $\alpha'=\alpha/\sin{45\degree}$. In Figure \ref{FIG8}(e) the hexagonal structure is shown on the left side and the pattern design on the right side. The structure consist of two hexagons shifted with respect to each other. Lattice parameters are \textit{a}=\textit{b}$\neq$\textit{c} and lattice angle $\alpha={120\degree}$. The pattern design consist of two completely different patterns. Pattern \textit{a} is similar to the well known hexagonal graphene-like pattern \cite{graphene}, whereas pattern \textit{b} has a rectangular array of apertures. Due to projection on the inclined surfaces the lattice parameters in the design are ${a'} =a/\sin{45\degree}$, ${c'} =c/\sin{45\degree}$ and ${b'} =b/\cos{\alpha'}$, where $\alpha'=\alpha/\sin{45\degree}$. The hexagonal mask has been realized in Cr on a Si wafer by means of a fabrication procedure described earlier in this article and is shown in Figure \ref{FIG9}. The red dashed line shows the $90\degree$ edge of a wafer. The top side on the picture is the (0001) crystal surface of hexagonal structure and the bottom side is a (1010) crystal plane. The 3D mask for this hexagonal structure shows clearly the flexibility of fabrication method to realise structures with independent patterns on inclined surfaces with an unprecedented out-of-plane alignment better than 5 nm. All described 3D structures except cubic are predicted to reveal sub-Bragg diffraction \cite{huis_bragg} which makes them interesting subject for optical study, moreover, simple cubic and hexagonal structures have been predicted to reveal 3D band gaps \cite{hex96,cubic}.

\begin{figure}[h!]
 \includegraphics[totalheight=12cm]{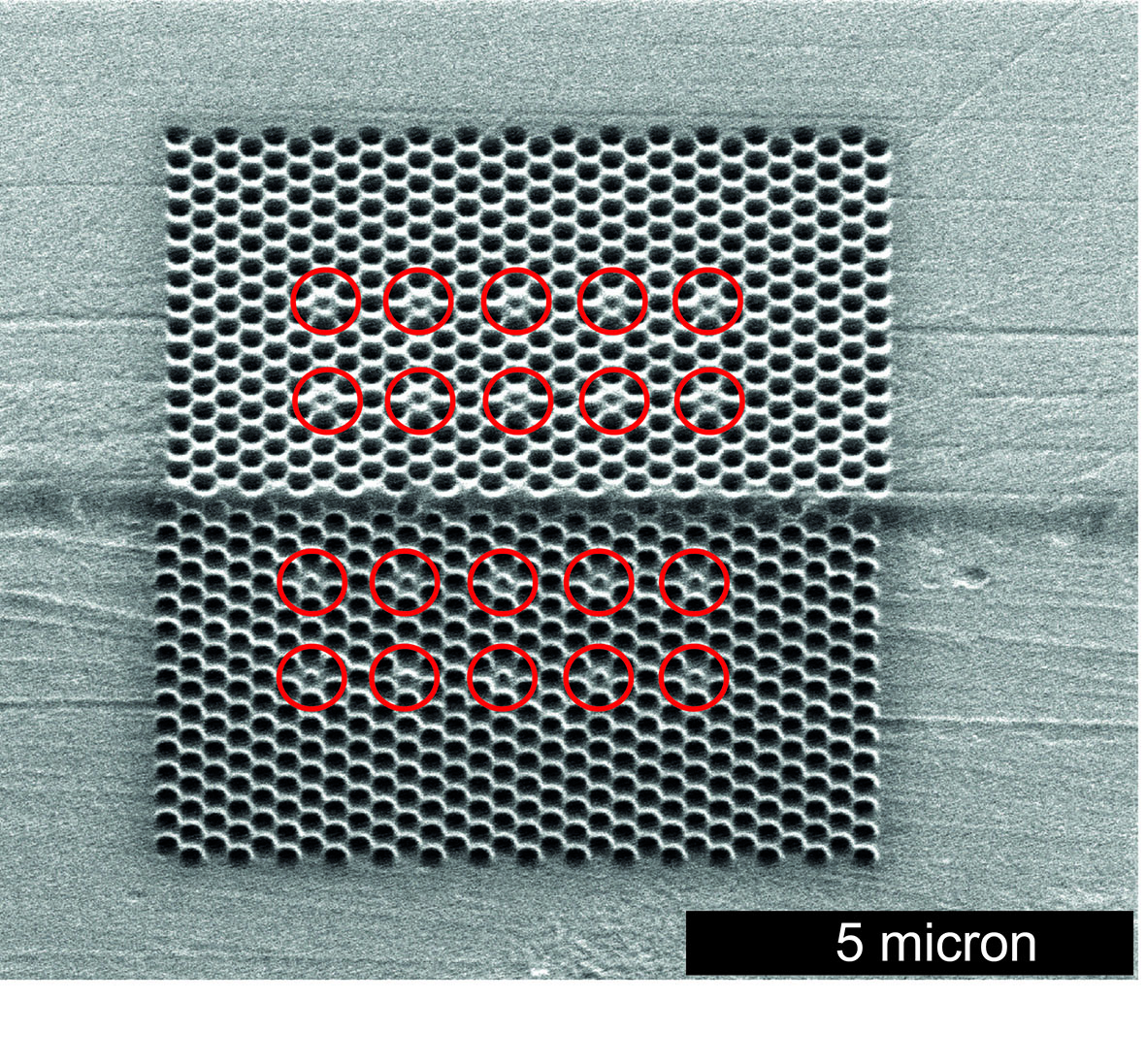}
 \caption{\label{FIG10} SEM image of a mask for an array of $2\times2\times4$ cavities in 3D photonic band gap crystal. Smaller pores that form cavities inside a structure are marked with red circles. (colour online) }%
 \end{figure}

 Besides different periodic structures it is possible to fabricate masks that yield non-periodic three-dimensional structures structures or periodic structures with controlled defects. An interesting example is a fabrication of a 3D photonic band gap crystal with a cavity inside. The geometry of cavity described in Reference \cite{cavity} can be straightforward realised with the presented fabrication method by  making two apertures for crossing pores on oblique planes smaller. Moreover, it is possible to realize an array of such cavities in a 3D photonic crystal. In Figure \ref{FIG10} the mask pattern for an array of $2\times2\times4$ cavities in 3D photonic crystal is shown. The resulting 3D array of band-gap cavities would represent the photonic version of the Anderson tight-binding model \cite{Anders} that may reveal intricate nanophotonic phase transitions for light. Another example is the fabrication of three-dimensional disordered or aperiodic structures. In this case patterns \textit{a} and \textit{b} can be arrays of randomly distributed apertures or incommensurable lattices \cite{rand15}. 
 
 Since patterns \textit{a} and \textit{b} are independent from each other, it is also possible to fabricate functionally different components on inclined surfaces. Figure 1 in Reference \cite{tjerk2011} shows an illustration of a suggested chip consisting of two integrated circuits on adjacent surfaces that are interconnected. Using the presented fabrication method it is possible to make interconnection between different integrated circuits with ensured alignment. In addition, it is possible to project optically functional device on one surface and electronic components on the other surface. Such architecture greatly increases density of components on chip, makes interconnections between them easier and gives a possibility to spatially separate optics and electronics on chip.
 
 The limitation for a number of possible structures comes from the subsequent silicon etching step. So far we assume that etching is always done with pores direction normal to the surface of the sample. Nevertheless there are examples where etching under an angle has been demonstrated \cite{noda_angle} which can further increase the number of feasible structures.

\section{Conclusions}

In summary, a novel method to fabricate a three-dimensional etch mask in one step with built-in alignment has been proposed. The out-of-plane alignment between structures on to oblique adjacent surfaces has been characterized by means of deviation from the designed structure. The out-of-plane alignment has been found to be better than 3.0 nm. A three-dimensional band gap photonic crystal with an inverse woodpile structure has been realized using a mask fabricated in a proposed way. The fabricated three-dimensional photonic crystal reveals a broad stop gap in optical reflectivity measurements. The masks designs for 3D nanostructures with five different Bravais lattices, namely cubic, tetragonal, orthorhombic, monoclinic, and hexagonal has been shown. The mask for 3D hexagonal structure and for a 3D array of cavities had been realised on a Si wafer.

\ack
The authors kindly thank Lyuba Amitonova for help in drawing 3D pictures, Oluwafemi Ojambati, Maryna Meretska, as well as Arie den Boef, Jo Finders (ASML), Robert van de Laar, Jeroen Bolk, Huub Ambrosius, Meint Smit (TU Eindhoven) and William Green (IBM) for fruitful discussions, Andreas Schultz for help with chemical cleaning of samples, Jorge Perez-Vizcaino for help with reflectivity measurements, Willem Tjerkstra and Johanna van den Broek for early contributions, and Allard Mosk for encouragements. This work was supported by ''Stirring of light!'' program of the Stiching voor Fundamenteel Onderzoek der Materie (FOM), which is financially supported by the Nederlandse Organisatie voor Wetenschappelijk Onderzoek (NWO) and by ``Stichting voor Technische Wetenschappen'' (STW).

\end{document}